\documentclass[aps,prd,superscriptaddress,nofootinbib,nobibnotes,showkeys,groupedaddress,10pt]{revtex4-2}
\usepackage{setspace}
\usepackage{color}
\usepackage{amsmath}
\usepackage{amsfonts}
\usepackage{verbatim}
\usepackage{amssymb}
\usepackage{graphicx,bm}
\usepackage{amsmath}
\usepackage{amssymb}
\usepackage{graphicx}
\usepackage[caption=false]{subfig}

\providecommand{\U}[1]{\protect\rule{.1in}{.1in}}

\newcommand{\be}{\begin{equation}}
\newcommand{\ee}{\end{equation}}

\newcommand{\mincir}{\raise
-3.truept\hbox{\rlap{\hbox{$\sim$}}\raise4.truept\hbox{$<$}\ }}
\newcommand{\magcir}{\raise
-3.truept\hbox{\rlap{\hbox{$\sim$}}\raise4.truept\hbox{$>$}\ }}

\begin{document}

\title{$f(R,\mathcal{G})$-cosmological dynamics in the FLRW background}
\author{Nikolaos Dimakis}
\email{nikolaos.dimakis@ufrontera.cl}
\affiliation{Departamento de Ciencias F\'{\i}sicas, Universidad de la Frontera, Casilla
54-D, 4811186 Temuco, Chile}
\author{Alex Giacomini}
\email{alexgiacomini@uach.cl }
\affiliation{Instituto de Ciencias F\'{\i}sicas y Matem\'{a}ticas, Universidad Austral de
Chile, Valdivia, Chile}
\author{Genly Leon}
\email{genly.leon@ucn.cl}
\affiliation{Departamento de Matem\'{a}ticas, Universidad Cat\'{o}lica del Norte, Avda.
Angamos 0610, Casilla 1280 Antofagasta, Chile}
\affiliation{Department of Mathematics, Faculty of Applied Sciences, Durban University of
Technology, Durban 4000, South Africa}
\author{Andronikos Paliathanasis}
\email{anpaliat@phys.uoa.gr}
\affiliation{Departamento de Matem\'{a}ticas, Universidad Cat\'{o}lica del Norte, Avda.
Angamos 0610, Casilla 1280 Antofagasta, Chile}
\affiliation{Department of Mathematics, Faculty of Applied Sciences, Durban University of
Technology, Durban 4000, South Africa}
\affiliation{National Institute for Theoretical and Computational Sciences (NITheCS),
South Africa}
\author{Ekaterina Pozdeeva}
\email{pozdeeva@www-hep.sinp.msu.ru}
\affiliation{Skobeltsyn Institute of Nuclear Physics, Lomonosov Moscow State University,
Leninskie Gory 1, Moscow 119991, Russia}
\author{Sergey Vernov}
\email{svernov@theory.sinp.msu.ru}
\affiliation{Skobeltsyn Institute of Nuclear Physics, Lomonosov Moscow State University,
Leninskie Gory 1, Moscow 119991, Russia}

\begin{abstract}
We examine the cosmological dynamics of Einstein-Gauss-Bonnet gravity models
in a four-dimensional spatially flat FLRW metric. These models are described
by $f\left( R,\mathcal{G}\right) =f\left( R+\mu \mathcal{G}\right) $ theory
of gravity. They are equivalent to models linear in the Ricci scalar $R$ and
in the Gauss-Bonnet scalar $\mathcal{G}$ with one nonminimally coupled
scalar field without kinetic term. We analyze the stability of the de Sitter
solutions and construct the phase space of the field equations to
investigate the cosmological evolution. We show that $f\left( R+\mu \mathcal{%
G}\right) $-theory provides a double inflationary epoch, this can be used to
unify the early-time and late-time acceleration phases of the universe.
Moreover, we discuss the initial value problem for theory to be
cosmologically viable. Finally, the effects of the cold dark matter in
cosmic evolution are discussed.
\end{abstract}

\maketitle

\section{Introduction}

The cosmological models with the Gauss-Bonnet term are motivated by string
theory~\cite%
{Antoniadis:1993jc,Kawai:1999pw,Cartier:2001is,Hwang:2005hb,Sami:2005zc,Calcagni:2005im,Tsujikawa:2006ph,Cognola:2006sp,Nojiri:2017ncd,Ketov:2022zhp,MohseniSadjadi:2023cjd}%
. The Gauss-Bonnet term affects the dynamics in higher-dimensional
gravitational models \cite%
{Lovelock:1971yv,Dotti:2007az,Pavluchenko:2008nh,Bogadi:2023moe,Pretel:2021gqq,Giacomini:2025hrd}%
. As is well known, the addition of the Gauss-Bonnet term to the General
Relativistic Lagrangian does not affect the evolution equations, since it is
a topological invariant in four dimensions.

To obtain a modified gravity model in four dimensions, one can multiply the
Gauss-Bonnet term by some nonconstant function of a scalar field \cite%
{Tsujikawa:2006ph,Sami:2005zc,Calcagni:2006ye,Koivisto:2006xf,Odintsov:2018zhw}%
. Models with the Ricci scalar $R$ and the Gauss-Bonnet term $\mathcal{G}$
multiplied by some functions of the scalar field are natural generalizations
of models with minimal coupling~\cite%
{vandeBruck:2015gjd,Mathew:2016anx,Pozdeeva:2021iwc}.

Inspired by $f\left( R\right) $-gravity \cite%
{Sotiriou:2008rp,DeFelice:2010aj,Nojiri:2009kx,Nojiri:2008nt,Amendola:2006kh}%
, natural extensions of the Gauss-Bonnet theory emerges the $f\left(
G\right) $-gravity~\cite{Nojiri:2005jg}, or the more general $f(R,\mathcal{G}%
)$-gravity~\cite%
{Bamba:2010wfw,delaCruz-Dombriz:2011oii,DeLaurentis:2015fea,Odintsov:2018nch,Elizalde:2020zcb}%
. In the latter case, the gravitational Action Integral can be written in an
equivalent form which is linear in $R$ and $\mathcal{G}$ that includes
scalar fields without kinetic \ terms~\cite%
{Nojiri:2005vv,Cognola:2006eg,Cognola:2006sp,Nojiri:2017ncd,Ketov:2022zhp,Vernov:2021hxo}%
.~The generic $f(R,\mathcal{G})$ can be linearized by two scalar fields as
we show in Appendix \ref{app1}. However, the two scalar fields are not
dynamically independent, which means that the real degrees of freedom can be
attributed to just one scalar field. It has been shown in Ref.~\cite%
{Vernov:2021hxo}, that $f\left( R+\mu \mathcal{G}\right) $ models, where $%
\mu $ is a constant, can be linearized by a single scalar field, which
describes all the dynamics.

A generic model with the Gauss-Bonnet term and a single scalar field can be
described by the following~Action Integral 
\begin{equation}
S=\!\int \!d^{4}x\sqrt{-g}\left[ U(\sigma )R-\frac{\beta }{2}g^{\mu \nu
}\partial _{\mu }\sigma \partial _{\nu }\sigma -V(\sigma )-F(\sigma )%
\mathcal{G}\right] ,  \label{action}
\end{equation}%
where $U(\sigma )$, $V(\sigma )$, and $F(\sigma )$ are double differentiable
functions of the scalar field $\sigma $. The Gauss-Bonnet term is given by 
\begin{equation}
\mathcal{G}=R^{2}-4R_{\mu \nu }R^{\mu \nu }+R_{\mu \nu \alpha \beta }R^{\mu
\nu \alpha \beta }.
\end{equation}%
At $\beta =1,$ we get\ standard scalar field models, whereas the case of $%
\beta =0$ corresponds to $f(R,\mathcal{G})$ models.

In this work, we are interested on the asymptotic cosmological dynamics
within the $f\left( R+\mu \mathcal{G}\right) $ theory and the stability
conditions for some exact solutions of special interest, such as the de
Sitter solution. There are different ways to analyze the stability of
solutions in cosmological models described by action~(\ref{action}). The
most popular is by introducing dimensionless variables and constructing
phase-space portraits~\cite%
{Lazkoz:2006pa,Curbelo:2005dh,Arefeva:2012sqa,Alhulaimi:2013sha,Skugoreva:2014gka,Oikonomou:2017ppp,Paliathanasis:2023nkb,Lohakare:2024ize,Duchaniya:2024vvc,Alho:2023xel,Bisabr:2023lqo,Samart:2021viu,Guarnizo:2020pkj,Aljaf:2019ilr,Carloni:2025kev,Carloni:2024ybx,DAgostino:2021vvv,Jarv:2021qpp,Papagiannopoulos:2025uix,Jarv:2025qgo}%
. This allows us to analyze not only stability of de Sitter solutions, but
also of scaling solutions. The corresponding phase-space analysis has been
widely studied in the case of four-dimensional Gauss-Bonnet theory with a
scalar field nonminimally coupled in \cite%
{Millano:2023czt,Millano:2023gkt,Paliathanasis:2024gwp}.

Another direction for the study for the stability analysis is to construct
the effective potential $V_{eff}$. The latter being a combination of
functions $U$, $V$, and $F$. In the case of $U(\sigma _{dS})>0,$ a stable de
Sitter solution correspond to a minimum of $V_{eff}$, whereas an unstable de
Sitter solutions correspond to a maximum of it~\cite%
{Pozdeeva:2019agu,Vernov:2021hxo}. In Refs.~\cite%
{Skugoreva:2014gka,Pozdeeva:2016cja}, both methods have been used to explore
dynamics of cosmological models with a nonminimally coupled scalar field,
which are described by the action (\ref{action}) with $\beta =1$ and $F(\phi
)=0$.

The structure of the paper is as follows. In Section \ref{sec2}, we
introduce the gravitational model, in which a scalar field, without a
kinetic term, is nonminimally coupled both to the Ricci scaler and to the
Gauss-Bonnet scalar. The theory of our consideration can describe the $%
f\left( R,\mathcal{G}\right) =f\left( R+\mu \mathcal{G}\right) $ gravity,
where $f$ is a nonlinear function. We show the equivalency between the two
theories by using the Lagrange multiplier approach. In Section III, we
recover the stability criteria of the de Sitter spacetime for the spatially
flat Friedmann-Lema\^{\i}tre-Robert\-son-Walker (FLRW) background geometry.
The phase-space analysis of the cosmological field equations is presented in
Section \ref{sec3}. We introduce dimensionless variables that allow the
Hubble function to change sign, providing new insights into cosmic evolution
and the theory's future attractors. In Section V, we examine the power-law
function $f\left( R,\mathcal{G}\right) \simeq \left( R+\mu \mathcal{G}%
\right) ^{\frac{n}{n-1}}$ both with and without a cosmological constant
term, and $f\left( R,\mathcal{G}\right) =R+\left( R+\mu \mathcal{G}\right) ^{%
\frac{n}{n-1}}$ theory. The cosmic evolution depends on the power $n$ of the
theory and the coefficient $\mu $. From the space-phase analysis it is found
that in order to avoid the existence of nonphysical solutions parameters $%
n,~\mu $ are constraint as $n>1$ or $\mu >\frac{1}{2M_{\mathrm{Pl}}^{2}}$.
The phase-space analysis reveals that most initial conditions lead to cosmic
collapse, a feature previously observed for the $f\left( R,\mathcal{G}%
\right) =R+F\left( \mathcal{G}\right) $ model in \cite%
{Papagiannopoulos:2025uix}. However, the theory favors future attractors
that describe singular universes. Additionally, we identify a subset of
initial conditions that pass through an unstable de Sitter expansion,
ultimately leading to another de Sitter solution as the future attractor.
This suggests that the theory can describe a double inflationary epoch or
establish a connection between early inflation and late-time acceleration.
The introduction of the cosmological constant plays a crucial role in
uncovering new physically viable behaviors within this modified gravity
framework. In Section \ref{sec4} we study the phase space in the presence of
cold dark matter. Finally, in Section \ref{sec5}, we summarize the results
of this study and present our conclusions.

\section{Einstein-Gauss-Bonnet Gravity and single-field models}

\label{sec2}

If the scalar field $\sigma $ has no kinetic term, $\beta =0$, then varying (%
\ref{action}) by $\sigma $, we obtain 
\begin{equation}
U_{,\sigma }R-F_{,\sigma }\mathcal{G}=V_{,\sigma },  \label{equsigma}
\end{equation}%
where the commas denote derivatives with respect to the scalar field, $%
A_{,\sigma }=\frac{dA}{d\sigma }$ for any function $A(\sigma )$. Using Eq.~(%
\ref{equsigma}), one can eliminate $\sigma $ from action (\ref{action}) and
get the corresponding $f(R,\mathcal{G})$ model.

For example, if we choose 
\begin{equation}
U(\sigma )=f_{,\sigma },\quad F(\sigma )=0,\quad V=\sigma f_{,\sigma }-f,
\end{equation}%
where $f(\sigma )$ is a nonlinear double differential function, then Eq.~(%
\ref{equsigma}) gives $\sigma =R$, and we obtain the $f(R)$ gravity model
with 
\begin{equation}
S_{\mathcal{F}}=\!\int \!d^{4}x\sqrt{-g}\,f(R).  \label{actionF}
\end{equation}%
The phase-space analysis of this type of models has been presented in \cite%
{Jarv:2025qgo}.

The choice of 
\begin{equation}
U(\sigma)=U_0,\quad F(\sigma)={\ } - f_{,\sigma}, \quad V=\sigma f_{,\sigma}
- f,
\end{equation}
where $U_0$ is a constant, gives us $\sigma=\mathcal{G}$ as a solution of
Eq.~(\ref{equsigma}). So, action (\ref{action}) transforms to 
\begin{equation}  \label{actionG}
S_{\mathcal{G}}=\!\int\! d^4 x\sqrt{-g}\left[U_0R +f(\mathcal{G})\right].
\end{equation}
The phase-space analysis of such type of models has been made in Ref.~\cite%
{Papagiannopoulos:2025uix}.

In this paper, we consider $f\left( R,\mathcal{G}\right) $ models that
correspond to nonconstant functions~$U(\sigma )$ and $F(\sigma )$. Namely,
we put 
\begin{equation}
U(\sigma )=f_{,\sigma },\quad F(\sigma )={}-\mu f_{,\sigma },\quad V=\sigma
f_{,\sigma }-f,
\end{equation}%
where $\mu =\mu _{0}/M_{\mathrm{Pl}}^{2}$, $\mu _{0}$ is a dimensionless
constant. We thus obtain $\sigma =R+\mu \mathcal{G}$, so action (\ref{action}%
) takes the form 
\begin{equation}
S_{R\,\mathcal{G}}=\!\int \!d^{4}x\sqrt{-g}\,f(R+\mu \mathcal{G}).
\label{actionFG}
\end{equation}

Note that the action 
\begin{equation}
S_{R\,\mathcal{G}}=\!\int \!d^{4}x\sqrt{-g}\,\left( U_{0}R+f_{1}(R+\mu 
\mathcal{G})\right)  \label{actionFGm}
\end{equation}%
belongs to the class of models of (\ref{actionFG}), because $\sqrt{-g}\,%
\mathcal{G}$ is a total derivative, so we can choose 
\begin{equation}
f(R+\mu \mathcal{G})=U_{0}(R+\mu \mathcal{G})+f_{1}(R+\mu \mathcal{G})
\end{equation}%
and get action (\ref{actionFGm}) from (\ref{actionFG}).

All functions (\ref{actionF}), (\ref{actionG}), and (\ref{actionFG}) are
particular cases of the action 
\begin{equation}
S_{R\,\mathcal{G}}=\!\int \!d^{4}x\sqrt{-g}\,\left( U_{0}R+f(c_{1}R+c_{2}%
\mathcal{G})\right) ,  \label{actionFGc1c2}
\end{equation}%
where $c_{1}$ and $c_{2}$ are constants.

\section{Stability of de Sitter solutions}

\label{sec2a}

In the spatially flat FLRW geometry with the line element%
\begin{equation}
ds^{2}={}-N^{2}\left( t\right) dt^{2}+a^{2}\left( t\right) \left(
dx^{2}+dy^{2}+dz^{2}\right) ,  \label{metric}
\end{equation}%
the Ricci scalar $R$ and the Gauss-Bonnet scalar $\mathcal{G}$ in terms of
the scale factor $a\left( t\right) $ and of the lapse function $N\left(
t\right) $ are as follows%
\begin{eqnarray}
R &=&6\left( 2H^{2}+\frac{1}{N}\dot{H}\right) ,  \label{f.02} \\
\mathcal{G} &=&24H^{2}\left( H^{2}+\frac{1}{N}\dot{H}\right) ,  \label{f.03}
\end{eqnarray}%
where the Hubble function is $H=\frac{\dot{a}}{N\,a}$ and the dots denote
derivatives with respect to $t$.

At $N=1$, when $t$ is the cosmic time, we derive the evolution equations: 
\begin{equation}
6H^{2}U+6HU_{,\sigma }\dot{\sigma}=V+24H^{3}F_{,\sigma }\dot{\sigma},
\label{Equ00}
\end{equation}%
\begin{equation}
2\left( U-4H\dot{F}\right) \dot{H}={}-\ddot{U}+H\dot{U}+4H^{2}\left( \ddot{F}%
-H\dot{F}\right) ,  \label{EquH}
\end{equation}%
\begin{equation}
V_{,\sigma }-6\left( \dot{H}+2H^{2}\right) U_{,\sigma }+24H^{2}F_{,\sigma
}\left( \dot{H}+H^{2}\right) =0.  \label{equphi}
\end{equation}

Equations (\ref{EquH}) and (\ref{equphi}) can be presented as the following
dynamical system~\cite{Vernov:2021hxo}: 
\begin{equation}
\begin{split}
\dot{\sigma}=& \zeta , \\
\dot{\zeta}=& \frac{1}{3\left( 4H^{2}F_{,\sigma }-U_{,\sigma }\right) ^{2}}%
\left\{ \left[ 12H^{2}\left( U_{,\sigma \sigma }F_{,\sigma }+U_{,\sigma
}F_{,\sigma \sigma }\right) -48F_{,\sigma }F_{,\sigma \sigma
}H^{4}-3U_{,\sigma \sigma }U_{,\sigma }\right] \zeta ^{2}\right. \\
+& \left. H\left[ 3B+4F_{,\sigma }V_{,\sigma }-6{U_{,\sigma }}^{2}\right]
\zeta -\frac{V^{2}}{U}\,X\right\} , \\
\dot{H}=& {}-\frac{V^{2}}{6\left( 4H^{2}F_{,\sigma }-U_{,\sigma }\right)
U^{2}}\,X\,,
\end{split}
\label{DynSYS}
\end{equation}%
where 
\begin{eqnarray}
X &=&\frac{U^{2}}{V^{2}}\left[ 24H^{4}F_{,\sigma }-12H^{2}U_{,\sigma
}+V_{,\sigma }\right] ,  \label{X} \\
B &=&3\left( 4H^{2}F_{,\sigma }-U_{,\sigma }\right) ^{2}.
\end{eqnarray}

A de Sitter solution corresponds to $\dot{\zeta}_{dS}=0$, $\zeta _{dS}=0$,
and $\dot{H}_{dS}=0$, so, $X(H_{dS},\sigma _{dS})=0$. Using Eq.~\eqref{Equ00}
and assuming $H_{dS}>0$, one gets 
\begin{equation}
H_{dS}=\sqrt{\frac{V(\sigma _{dS})}{6U(\sigma _{dS})}\,},  \label{HDS}
\end{equation}%
hence, 
\begin{equation}
X(H_{dS},\sigma _{dS})=\frac{2}{3}F_{,\sigma }(\sigma _{dS})-\frac{2U(\sigma
_{dS})U_{,\sigma }(\sigma _{dS})}{V(\sigma _{dS})}+\frac{U^{2}(\sigma
_{dS})V_{,\sigma }(\sigma _{dS})}{V^{2}(\sigma _{dS})}=V_{eff,\sigma
}(\sigma _{dS})=0,  \label{XdS}
\end{equation}%
where the effective potential is 
\begin{equation}
V_{eff}=\frac{2}{3}F{}-\frac{U^{2}}{V}={}-\frac{2\mu }{3}\,f_{,\sigma }-%
\frac{f_{,\sigma }^{2}}{\sigma f_{,\sigma }-f}\,.  \label{Veff}
\end{equation}%

In the case of $U(\sigma _{dS})>0$, stable de Sitter solutions correspond to 
$V_{eff,\sigma \sigma }(\sigma _{dS})>0$, whereas unstable de Sitter
solutions correspond to $V_{eff,\sigma \sigma }(\sigma _{dS})<0$, the proof
is given in Ref.~\cite{Vernov:2021hxo}. Note that the effective potential
can be useful for the consideration of slow-roll inflation in models
described by the action~(\ref{action}) with $\beta =1$, see Refs.~\cite%
{Pozdeeva:2020apf,Pozdeeva:2024ihc,Pozdeeva:2025ied}.

In the spatially flat FLRW metric, we now set $N=1.$ Accordingly to the
definition of $\sigma $ for the $f(R+\mu \mathcal{G})$ models, we get 
\begin{equation}
\sigma =6\left( 2H^{2}+\dot{H}\right) +24\mu \,H^{2}\left( H^{2}+\dot{H}%
\right) ,  \label{sigmaFLRW}
\end{equation}%
so Eq.~(\ref{Equ00}) takes the following form: 
\begin{equation}
\begin{split}
& 36H\left( 4\mu H^{2}+1\right) \left( 4H^{2}F_{,\sigma }-U_{,\sigma
}\right) \ddot{H}+288\mu H^{2}\left( 4H^{2}F_{,\sigma }-U_{,\sigma }\right) {%
\dot{H}}^{2} \\
& {}+144H^{2}\left( 4\mu H^{2}+1\right) \left( 4H^{2}F_{,\sigma }-U_{,\sigma
}\right) \dot{H}=6H^{2}U-V\,.
\end{split}
\label{Equ00H}
\end{equation}

One can see that formula (\ref{HDS}) is a trivial consequence of Eq.~(\ref%
{Equ00H}).

It follows from Eq.~(\ref{sigmaFLRW}) that 
\begin{equation}  \label{sigmaepsilon1}
\sigma=12H^2\left(1-\frac12
\epsilon_1+2\mu\,H^{2}\left(1-\epsilon_1\right)\right),\qquad \epsilon_1={}-%
\frac{\dot{H}}{H^2},
\end{equation}
so at $\mu>0$, an accelerating expansion of the Universe requires $\sigma>0$.

\section{Evolution equations in terms of dimensionless variables}

\label{sec3}

For the phase-space analysis of $f(R+\mu \mathcal{G})$ models, it is
suitable to introduce a new scalar field~$\phi $ 
\begin{equation}
\phi =U\left( \sigma \right) ,  \label{Uphi}
\end{equation}%
therefore, $F\left( \sigma \right) ={}-\mu \phi $ and the effective
potential becomes 
\begin{equation}
V_{eff}=\frac{2}{3}F{}-\frac{U^{2}}{V}={}-\frac{2}{3}\mu \phi -\frac{\phi
^{2}}{V(\phi )}.
\end{equation}%
To obtain $V(\phi )$ one should invert Eq.~(\ref{Uphi}) and solve with
respect to $\sigma $.

The evolution equations for the spatially flat FLRW metric (\ref{metric})
follow from the variation of the point-like Lagrangian, 
\begin{equation}
L_{f\left( R,G\right) }^{A}\left( N,a,\dot{a},\phi ,\dot{\phi}\right) ={}-%
\frac{6}{N}\phi a\dot{a}^{2}-\frac{6}{N}a^{2}\dot{a}\dot{\phi}-\frac{8\mu }{%
N^{3}}\dot{\phi}\dot{a}^{3}-Na^{3}V\left( \phi \right) .  \label{ac.10}
\end{equation}%
and have the following form 
\begin{equation}
6\phi H^{2}+6H\frac{\dot{\phi}}{N}+24H^{3}\mu \frac{\dot{\phi}}{N}=V\left(
\phi \right) ,  \label{ac.11}
\end{equation}%
\begin{eqnarray}
0 &=&\frac{4}{N}\left( \frac{d}{dt}\left( \phi H\right) +4\mu H^{3}\dot{\phi}%
\right) +\frac{2}{N^{2}}\left( 8\mu H\dot{H}\dot{\phi}+\left( 1+4\mu
H^{2}\right) \ddot{\phi}\right)  \notag \\
&-&\frac{2}{N^{3}}\left( 1+4\mu H^{2}\right) \dot{N}\dot{\phi}+6\phi
H^{2}-V\left( \phi \right) ,  \label{ac.12}
\end{eqnarray}%
\begin{equation}
V_{,\phi }=6\left( 2H^{2}+\frac{1}{N}\dot{H}\right) +24\mu H^{2}\left( H^{2}+%
\frac{1}{N}\dot{H}\right) \text{.}  \label{ac.13}
\end{equation}

More general $f\left( R,\mathcal{G}\right) $-gravity models can be written
in the equivalent form using two scalar fields. The evolution equations for
the FLRW metric in such models are given in Appendix~\ref{app1}.

We proceed with the analysis of the phase space for the latter set of
nonlinear differential equations. Specifically, we examine the existence of
equilibrium points. Each equilibrium point describes a specific era in the
cosmological history. The stability properties of the equilibrium points
give information about cosmic evolution. In the following, without loss of
generality, we assume the lapse function $N=1$.

From the constraint equation (\ref{Equ00}), we note that the Hubble function 
$H$ can change its sign and take the value zero. Thus, we study the phase
space dynamics by introducing appropriate dimensionless variables where the
Hubble function can change its sign smoothly.

We consider the following dimensionless variables \cite%
{Millano:2023czt,Millano:2023gkt,Paliathanasis:2024gwp} 
\begin{equation}
x=\frac{8\dot{\phi}}{\phi \sqrt{M_{\mathrm{Pl}}^{2}+H^{2}}},\quad y={}-\frac{%
V\left( \phi \right) }{\phi \left( M_{\mathrm{Pl}}^{2}+H^{2}\right) },\quad
\eta =\frac{H}{\sqrt{M_{\mathrm{Pl}}^{2}+H^{2}}},  \label{xyeta}
\end{equation}%
\begin{equation}
\lambda =\phi \frac{V_{,\phi }}{V},\quad \tau =\int \sqrt{M_{\mathrm{Pl}%
}^{2}+H^{2}}dt,  \label{lambdatau}
\end{equation}%
in which we assume $\tau $ to be a new independent variable (a parametric
time), whereas $x,~y,~\eta $ and $\lambda $ to be new dependent variables.

It follows from Eqs.~(\ref{ac.13}) and (\ref{xyeta}) that 
\begin{equation}  \label{Heta}
H=\frac{M_{\mathrm{Pl}}\,\eta}{\sqrt{1-\eta^2}}\,,
\end{equation}
\begin{equation}
\dot{H}=\frac{M_{\mathrm{Pl}}^2\left(
24\mu_0\eta^4-12\eta^4-\eta^2\lambda\,y+12\eta^2+y\lambda\right)}{%
6\left(\eta^2-1\right)\left( (4\mu_0-1)\eta^2+1\right)}\,.
\end{equation}

In terms of the dimensionless variables, Eq.~(\ref{ac.11}) leads to the
algebraic constraint 
\begin{equation}
y=-6\eta ^{2}-\frac{3}{4}\,\eta \,x\left( 1+\frac{4\mu _{0}\eta ^{2}}{1-\eta
^{2}}\right) ,  \label{ac.19}
\end{equation}%
and the deceleration parameter reads%
\begin{equation}
q=\frac{\left( 1-\lambda \right) \left( 1-\eta ^{2}\right) }{\left( 1-\left(
1-4\mu _{0}\right) \eta ^{2}\right) }-\frac{\lambda }{8}\frac{x}{\eta }\,.
\label{ac.20}
\end{equation}

We also get, 
\begin{equation}
\frac{d\eta}{d\tau}=\left(1-\eta^2\right)^2\frac{\dot{H}}{M_{\mathrm{Pl}}^2}%
= {}-\frac{\left( 1-\eta ^{2}\right) \left( \lambda y\left( 1-\eta
^{2}\right) +12\eta ^{2}\left( 1-\left( 1-2\mu_0 \right) \eta ^{2}\right)
\right) }{6\left( 1-\left( 1-4\mu_0 \right) \eta^{2}\right) }.  \label{ac.17}
\end{equation}

Using 
\begin{equation}
\dot{\phi}=\frac{1}{8}x\phi \,\sqrt{H^{2}+M_{\mathrm{Pl}}^{2}},
\end{equation}%
\begin{equation}
\ddot{\phi}={}-\frac{48\mu _{0}\eta ^{4}+2\mu _{0}\eta \lambda xy+12\eta
^{4}+3\eta ^{3}x+4\eta ^{2}\lambda y+24\mu _{0}\eta ^{2}y-6\eta ^{2}y-12\eta
^{2}-3\eta x-4\lambda y+6y}{12\left( 4\mu _{0}\eta ^{2}-\eta ^{2}+1\right)
^{2}}\phi ,
\end{equation}%
and Eq.~(\ref{ac.12}), we obtain 
\begin{eqnarray}
\frac{dx}{d\tau } &=&{}-\frac{1}{8}x^{2}+\frac{\eta x\left( \lambda y\left(
1-\eta ^{2}\right) +12\eta ^{2}\left( 1-\left( 1-2\mu _{0}\right) \eta
^{2}\right) \right) }{6\left( 1-\left( 1-4\mu _{0}\right) \eta ^{2}\right) }
\notag \\
&&{}-\frac{2\left( 1-\eta ^{2}\right) \left( y\left( 6-4\lambda -2\lambda
\mu _{0}x\eta -\left( 6-4\lambda -24\mu _{0}\right) \eta ^{2}\right) \right) 
}{3\left( 1-\left( 1-4\mu _{0}\right) \eta ^{2}\right) ^{2}}  \notag \\
&&-\frac{2\left( 1-\eta ^{2}\right) \eta \left( 4\left( 1+4\mu _{0}\right)
\eta ^{3}-4\eta +x\left( 1-\eta ^{2}\right) \right) }{\left( 1-\left( 1-4\mu
_{0}\right) \eta ^{2}\right) ^{2}}\,.  \label{ac.15}
\end{eqnarray}

Similarly, we deduce 
\begin{equation}
\begin{split}
\frac{dy}{d\tau }& =\frac{y\left( 8\lambda y\eta \left( 1-\eta ^{2}\right)
-3\left( 1-\lambda \right) x\left( 1-\left( 1-4\mu _{0}\right) \eta
^{2}\right) +96\eta ^{3}\left( 1-\left( 1-2\mu _{0}\right) \eta ^{2}\right)
\right) }{24\left( 1-\left( 1-4\mu _{0}\right) \eta ^{2}\right) } \\
& =\frac{\left( \lambda -1\right) yx}{8}+\frac{\left( 24\mu _{0}\eta
^{4}-12\eta ^{4}-\eta ^{2}\lambda \,y+12\eta ^{2}+y\lambda \right) \eta \,y}{%
3\left( 1+\left( 4\mu _{0}-1\right) \eta ^{2}\right) }\,,
\end{split}
\label{ac.16}
\end{equation}%
and 
\begin{equation}
\frac{d\lambda }{d\tau }=\frac{\lambda }{8}x\left( 1-\lambda +\lambda \Gamma
\left( \lambda \right) \right) ~\text{where}\quad ~\Gamma \left( \lambda
\right) =\frac{V_{,\phi \phi }V}{V_{,\phi }^{2}}.  \label{ac.18}
\end{equation}

By definition, the dynamical variable $\eta $ is governed by the constraint $%
\left\vert \eta \right\vert \leqslant 1$. However, the rest of the dynamical
variables are not subject to constraints and assume values in whole set of
the real numbers.

{Due to the algebraic constraint equation (\ref{ac.19}), the order of the
dynamical system is reduced by one. We substitute the dynamical variable $y$
from (\ref{ac.19}) in the rest of the field equations and obtain a
third-order dynamical system. It is better to eliminate $y$ instead of $x$,
because at $\eta =0$, Eq.~(\ref{ac.19}) gives $y=0$, whereas there is no
restriction on the values of $x$. We proceed with the derivation of the
equilibrium points for this dynamical system for specific functional forms
of the potential.}

\section{Phase-space analysis of $f(R+\protect\mu \mathcal{G})$ models}

\subsection{Examples of $f(R+\protect\mu \mathcal{G})$ models}

Let us consider 
\begin{equation}
f(R+\mu \mathcal{G})=U_{0}(R+\mu \mathcal{G})+\alpha (R+\mu \mathcal{G}%
)^{n/(n-1)}-\Lambda ,  \label{frG}
\end{equation}%
where $U_{0}$, $\alpha $, $\Lambda $, and $n\neq 1$ are constants. At $%
\alpha =0$ or $n=0$, we retrieve the General Relativistic model with a
cosmological constant. We explore the case of modified gravity, when $\alpha
\neq 0$ and $n\neq 0$.

For the function 
\begin{equation*}
f(\sigma)=U_0\sigma+\alpha \sigma^{n/(n-1)}-\Lambda,
\end{equation*}
we get 
\begin{equation}  \label{phisigma}
\phi=f_{,\sigma}=U_0+\frac{\alpha n}{n-1}\sigma^{1/(n-1)},
\end{equation}
\begin{equation}
V=\sigma f_{,\sigma}-f=\frac{\alpha}{n-1}\sigma^{n/(n-1)}+\Lambda=V_0\left(%
\phi-U_0\right)^n+\Lambda,
\end{equation}
where 
\begin{equation}  \label{V0}
V_0=\frac{\alpha}{n-1}\left(\frac{n-1}{\alpha n}\right)^n.
\end{equation}

The effective potential is the 
\begin{equation}
V_{eff}={}-\frac{2}{3}\,\mu \left( U_{0}+\frac{\alpha n\,{\sigma }^{n/\left(
n-1\right) }}{n-1}\right) -\frac{\left( \alpha \,n{\sigma }^{{n}/{(n-1)}%
}+(n-1)U_{0}\,\sigma \right) ^{2}}{\left( n-1\right) {\sigma }^{2}\left(
\Lambda (n-1)+\alpha \,{\sigma }^{n/(n-1)}\right) }\,.
\end{equation}

In the following subsections, we consider a few special cases of the
potential $V(\phi)$.

\subsection{Power-law potential $V\left( \protect\phi \right) =V_{0}\protect%
\phi^{n}$}

\subsubsection{de Sitter solutions}

In the case of $U_0=0$ and $\Lambda=0$, we get the power-law potential $%
V\left( \phi \right) =V_{0}\phi^{n}$ and 
\begin{equation}  \label{fn}
f\left(\sigma\right) =\alpha\,\sigma^{\frac{n}{n-1}}.
\end{equation}

The effective potential is 
\begin{equation}
V_{eff}(\sigma )={}-\frac{n\alpha }{3(n-1)}\left( 3n\sigma
^{-(n-2)/(n-1)}+2\mu \sigma ^{1/(n-1)}\right) \,,  \label{Veff_fN}
\end{equation}%
therefore, 
\begin{equation}
\sigma _{dS}=\frac{3(n-2)n}{2\mu }\,.  \label{phidS}
\end{equation}

The effective potential has been derived from the third-order system (\ref%
{DynSYS}). In terms of $H$ and its derivatives, this system is a
differential consequence of Eq.~(\ref{Equ00H}). For de Sitter solution, Eq.~(%
\ref{Equ00H}) gives the following solution 
\begin{equation}
H_{dS}^{2}=\frac{V(\sigma _{dS})}{6U(\sigma _{dS})}=\frac{(n-2)M_{\mathrm{Pl}%
}^{2}}{4\mu _{0}}\,.  \label{Equ00dS}
\end{equation}%
The solution $H_{dS}$ exists at $(n-2)\mu _{0}>0$. The corresponding value
of $\eta _{dS}$ is 
\begin{equation}
\eta _{dS}=\sqrt{\frac{2-n}{2-n-4\mu _{0}}},
\end{equation}%
where we use $H_{dS}>0$.

At a de Sitter point, we get 
\begin{equation}  \label{ydS}
y_{dS}={}-\frac{\sigma_{dS}}{n \left(M_{\mathrm{Pl}}^2+H^{2}\right)}={}-%
\frac {3\left( n-2 \right)M_{\mathrm{Pl}}^2}{2\mu_0\, \left(M_{\mathrm{Pl}%
}^2+H^{2}\right)}\,.
\end{equation}

Using Eq.~(\ref{ac.19}), we obtain 
\begin{equation}
x=\frac {4 \left( 6\,{\eta}^{2}+y \right) \left( {\eta}^{2}-1\right)}{3\eta
\left( 4 \mu_0{\eta}^{2}-{\eta}^{2}+1\right)}={}- \frac {4 M_{\mathrm{Pl}%
}^2\left(6H^2\phi-V\right)}{3H\phi\left(4\mu_0 H^2+M_{\mathrm{Pl}}^2 \right)%
\sqrt{M_{\mathrm{Pl}}^2+ H^2\,}}\,.
\end{equation}

So, 
\begin{equation}  \label{xdS}
x_{dS}=0\,.
\end{equation}

Note that by definition, 
\begin{equation}  \label{xH}
x=\frac{8(\dot{R}+\mu\dot{\mathcal{G}})}{(n-1)(R+\mu{\mathcal{G}})\sqrt{M_{%
\mathrm{Pl}}^2+H^2}}=\frac{8\left[8\mu H{\dot{H}}^2+4H\left(4\mu
H^{2}+1\right){\dot{H}}+ \left(4\mu H^{2}+1\right)\ddot{H}\right]} {(n-1)%
\left[\left(4\mu H^{2}+1\right)\dot{H} +4\mu H^{4}+2H^{2}\right]\sqrt{M_{%
\mathrm{Pl}}^2+H^2}},
\end{equation}
so any de Sitter solution corresponds to $x=0$.

The search of fixed points of the equations in terms of dimensionless
variable allows to get this de Sitter solution as well, it is the point $%
P_{4}^{+}$ (see below).

Let us make the following remark. There is another way to obtain values of
the Hubble parameter from $\sigma_{dS}$. Solving equation 
\begin{equation}  \label{sigmadS}
12H_{dS}^2+24\mu H_{dS}^4=\sigma_{dS}=\frac{3(n-2)n}{2\mu }\,,
\end{equation}
one obtains two solutions: 
\begin{equation}  \label{HsigmadS}
\tilde{H}_{dS}^2=\frac{(n-2) M_{\mathrm{Pl}}^2}{4\mu_0},\qquad \hat{H}%
_{dS}^2={}-\frac{n M_{\mathrm{Pl}}^2}{4\mu_0},
\end{equation}
but only $\tilde{H}_{dS}$ is a de Sitter solution of the model considered.

\subsubsection{Solutions in dimensionless coordinates}

Formula (\ref{lambdatau}) implies that a monomial potential $V(\phi )$
corresponds to a constant $\lambda =n$. Therefore, we examine the
phase-space for the two-dimensional space with dynamical variables $x$ and $%
\eta $.

It follows that the field equations possess nine equilibrium points~$%
P=\left( x\left( P\right) ,\eta \left( P\right) \right) $. The physical
properties and the stability of the asymptotic solutions near to the
stationary points are described below:

\begin{itemize}
\item The fixed point 
\begin{equation*}
P_{0}=\left( 0,0\right) ,
\end{equation*}%
describes the Minkowski spacetime, where the scale factor is a constant. The
eigenvalues of the linearized system around the point are zero. Hence, the
point is a center and the solution is unstable.

\item The fixed points 
\begin{equation*}
P_{1}^{\pm }=\left( 0,\pm 1\right) ,
\end{equation*}%
describe scaling solutions where the Gauss-Bonnet term dominates, that is $%
q\left( P_{1}^{\pm }\right) =0$. With the scaling solutions we refer to
power-law scale factors in which the background geometry admits a homothetic
symmetry given by a scaling vector field. The corresponding eigenvalues of
the linearized system are calculated for point $\left\{ \pm 4,\pm 2\right\} $%
, from where we infer that $P_{1}^{+}$ is always a source, and point $%
P_{1}^{-} $ is always an attractor.

\item The fixed points 
\begin{equation*}
P_{2}^{\pm }=\left( \pm \frac{32}{1+3n},\pm 1\right)
\end{equation*}%
exist for $n\neq -\frac{1}{3}$. The points describe scaling solutions with $%
q\left( P_{2}^{\pm }\right) =-\frac{4n}{1+3n}$, that is, acceleration is
occurred for $n<-\frac{1}{3}$ and $n>0$. For large values of $\left\vert
n\right\vert $, it follows that $q\left( P_{2}^{\pm }\right) \rightarrow -%
\frac{4}{3}$. The eigenvalues of the linearized system near to the
stationary points are $\left\{ \frac{2\left( 1-n\right) }{1+3n},-4\right\} $
for point $P_{2}^{+}$ and $\left\{ -\frac{2\left( 1-n\right) }{1+3n}%
,4\right\} $ for point $P_{2}^{-}$.\ Hence, for $n<-\frac{1}{3}$or~$n>0,$
point $P_{2}^{+}$ is an attractor and $P_{2}^{-}$ is a source. Nevertheless,
for $n$ within the region $-\frac{1}{3}<n<0$, the equilibrium points are
saddle points.

\item The fixed points 
\begin{equation*}
P_{3}^{\pm }=\frac{1}{\sqrt{1-2\mu _{0}}}\left( \pm 8,\pm 1\right)
\end{equation*}%
are real and physically accepted for $\mu _{0}<\frac{1}{2}$. For these
points we calculate that $q\left( P_{3}^{\pm }\right) =-1.$ Someone may
infer that the asymptotic solutions describe the de Sitter universe, but on
the other hand, $\eta ^{2}=1$. Thus, these points are not physically
accepted. The eigenvalues of the linearized system around the stationary
points are $\left\{ -\frac{1-n}{\sqrt{1-2\mu _{0}}},-\frac{4}{\sqrt{1-2\mu
_{0}}}\right\} $ for point $P_{3}^{+}$, and $\left\{ \frac{1-n}{\sqrt{1-2\mu
_{0}}},\frac{4}{\sqrt{1-2\mu _{0}}}\right\} $ for point $P_{3}^{-}$. Hence,
for parameter $n<1$, point $P_{3}^{+}$ is an attractor and $P_{3}^{-}$ is a
source. Otherwise, for $n\geq 1$ the points are saddle points. In order to
avoid these points, parameter $n$ should be greater than one, $n>1$. On the
other hand, points $P_{3}^{\pm }$ do not exist at $\mu _{0}>\frac{1}{2}$.

\item The fixed points 
\begin{equation*}
P_{4}^{\pm }=\left( 0,\pm \sqrt{\frac{2-n}{2-n-4\mu _{0}}}\right) ,
\end{equation*}%
are real and physically accepted for $\left\{ \mu _{0}<0:n<2\right\} $ or $%
\left\{ \mu _{0}>0:2<n~\right\} $.

The asymptotic solutions have deceleration parameter $q\left( P_{4}^{\pm
}\right) =-1$, {so, $P_{4}^{+}$ describes de Sitter universe, whereas $%
P_{4}^{-}$ describes universe with a constant negative Hubble parameter.}
The analysis of the eigenvalues gives that point $P_{4}^{+}$ is an attractor
when $\left\{ \mu _{0}<0,~1<n<2\text{ }\right\} {,~}$while point $P_{4}^{-}$
is a source.
\end{itemize}

\subsubsection{Compactified variables}

In order to examine the existence of stationary points for very large values
of the parameter $x$, we introduce the new dependent and independent
variables%
\begin{equation}
x=\frac{X}{\sqrt{1-X^{2}}},~\sqrt{1-X^{2}}dT=d\tau ,  \label{ac.22}
\end{equation}%
where the parameter $X$ is constrained as $\left\vert X\right\vert \leq 1$.

The two-dimensional dynamical system takes the form 
\begin{eqnarray}
\frac{dX}{dT} &=&F_{1}\left( X,\eta ; n \right) , \\
\frac{d\eta }{dT} &=&F_{2}\left( X,\eta ; n \right) .
\end{eqnarray}

There are six equilibrium points which solve the algebraic equations $%
X^{2}=1 $ and $\eta \left( \eta ^{2}-1\right) =0$.

The phase-space analysis at the infinity regime is summarized in the
following lines.

\begin{itemize}
\item The fixed points 
\begin{equation*}
P_{0}^{\left( \infty \right) \pm }=\left( \pm 1,0\right) ,
\end{equation*}%
describe Minkowski solutions, with eigenvalues $\left\{ \pm \frac{n}{8},\pm 
\frac{1}{4}\right\} $, from where infer that for $n>0$, $P_{0}^{\left(
\infty \right) +}$ is a source point, while $P_{0}^{\left( \infty \right) -}$
is an attractor.

\item The fixed points 
\begin{equation*}
P_{1}^{\left( \infty \right) \pm }=\left( \pm 1,1\right) .
\end{equation*}%
correspond to universes with $q\left( P_{1}^{\left( \infty \right) \pm
}\right) =-\frac{n}{8}\infty $. Hence, for $n>0$, the points describe Big\
Rip singularities, however for $n<0$, they describe Minkowski spacetimes.
Because for $n<0$, $q\left( P_{1}^{\left( \infty \right) \pm }\right) >0$,
but $\eta >0$, which means that necessary the asymptotic solution is that of
the Minkowski spacetime. The eigenvalues of the linearized system are
determined $\left\{ \mp \frac{n}{4},\pm \frac{1}{4}\left( 1+3n\right)
\right\} $. Therefore only point $P_{1}^{\left( \infty \right) -}$ can be an
attractor, for $-\frac{1}{3}<n<0$.

\item The fixed points 
\begin{equation*}
P_{-1}^{\left( \infty \right) \pm }=\left( \pm 1,-1\right)
\end{equation*}%
correspond\ to universes with $q\left( P_{-1}^{\left( \infty \right) \pm
}\right) =\frac{n}{8}\infty $. Hence for $n>0$ the points describe Big
Crunch singularities, however for $n<0$, they describe Minkowski spacetimes,
because $\eta <0$. The eigenvalues are calculated~$\left\{ \mp \frac{n}{4}%
,\pm \frac{1}{4}\left( 1+3n\right) \right\} $. Therefore only point $%
P_{-1}^{\left( \infty \right) -}$ can be an attractor, for $-\frac{1}{3}<n<0$%
.
\end{itemize}

The results of the phase-space analysis for the exponential potential are
summarized in Table~\ref{tab2}.

\begin{figure}[tbph]
\centering\includegraphics[width=1\textwidth]{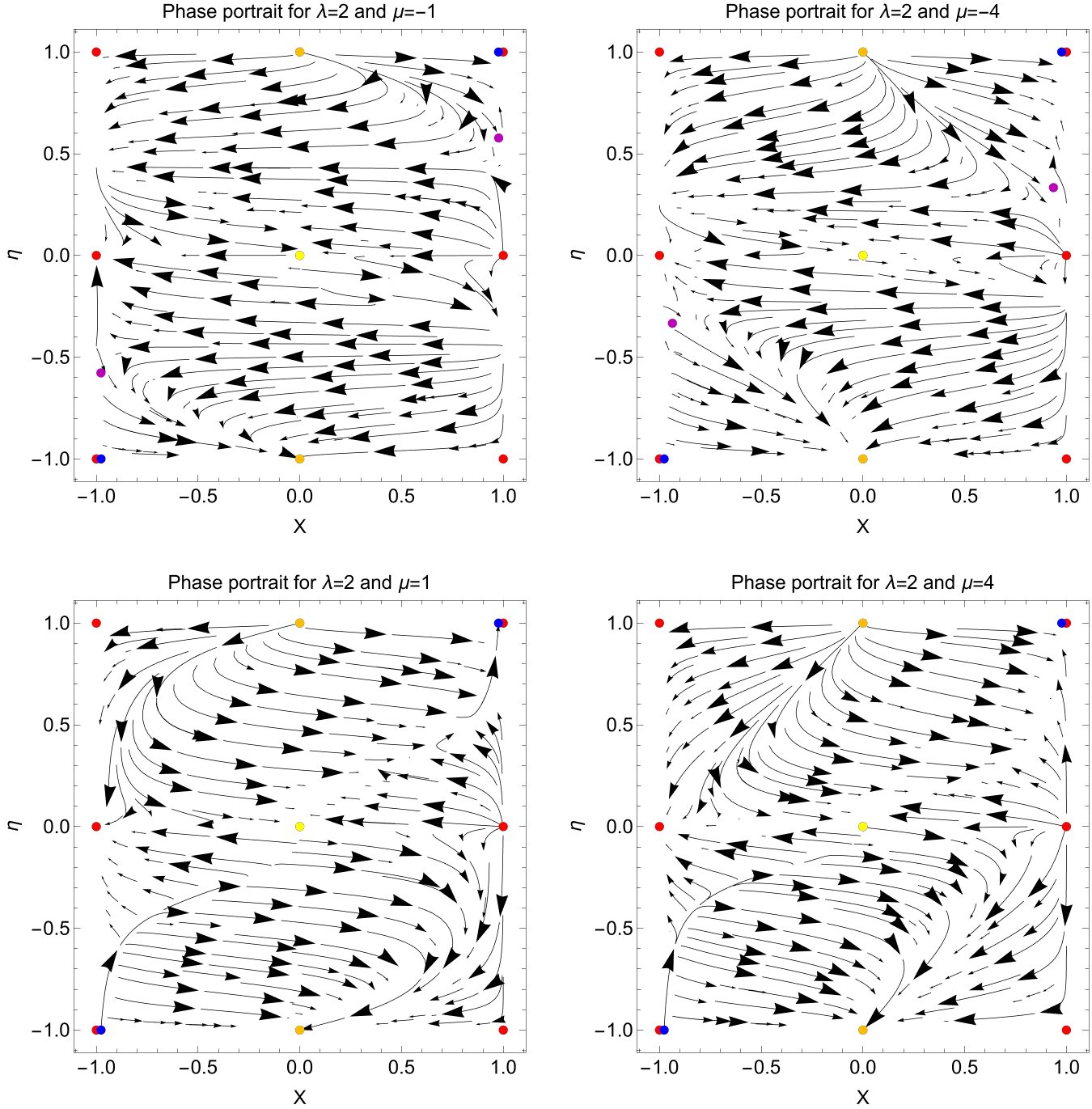}
\caption{Phase-space portrait for the two-dimensional dynamical system (%
\protect\ref{ac.15}), (\protect\ref{ac.17}) in the compactified variables $X$
and $\protect\eta ~$ for $n=2$ and various values of the parameter~$\protect%
\mu _{0}$. Equilibrium points at the infinity are marked with red. Point $%
P_{0}$ is marked with yellow, orange points are $P_{1}^{\pm }$, blue points
are $P_{2}^{\pm }$, and purple points are $P_{3}^{\pm }$. }
\label{fig1}
\end{figure}

\begin{figure}[tbph]
\centering\includegraphics[width=1\textwidth]{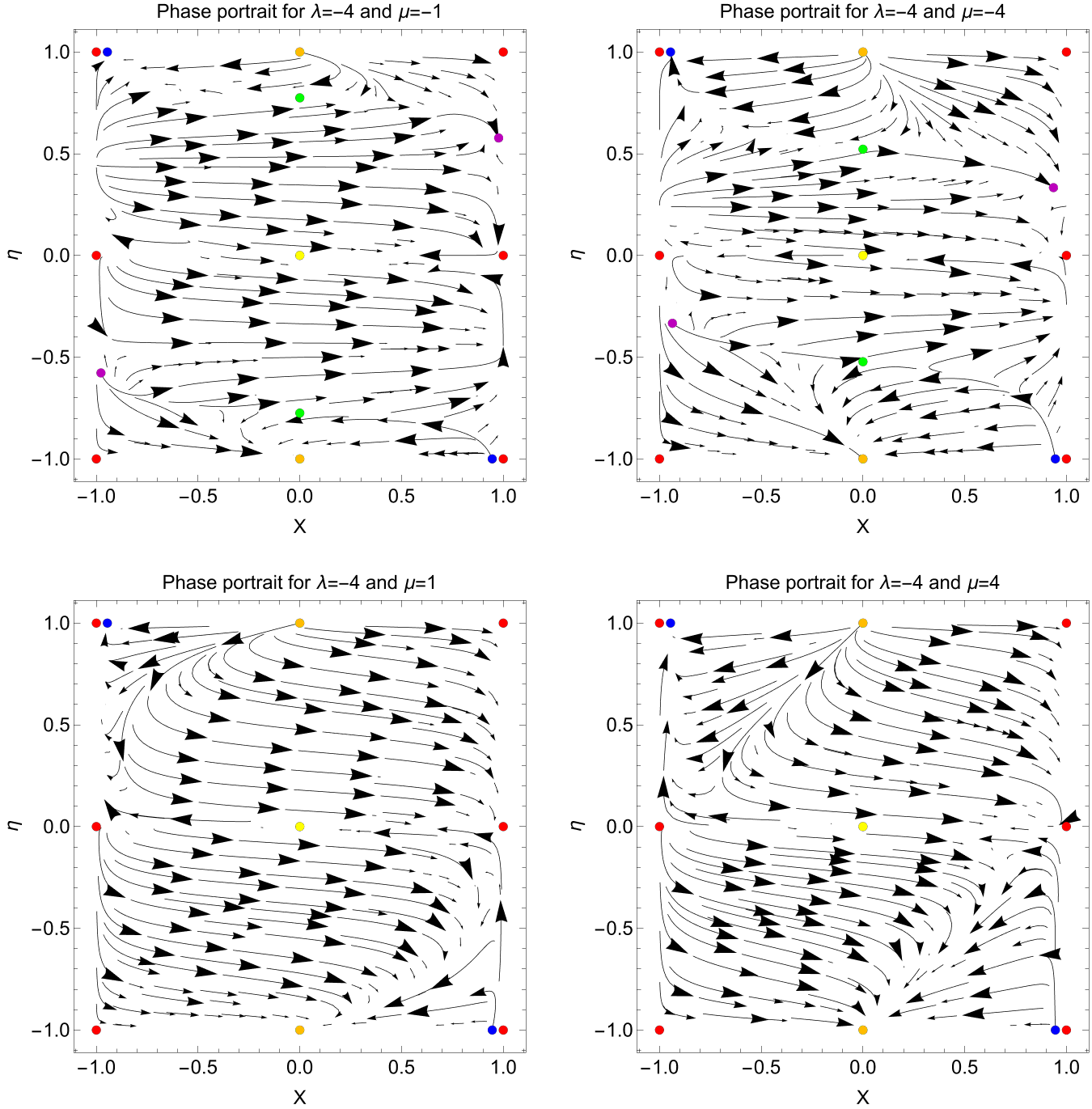}
\caption{Phase-space portrait for the two-dimensional dynamical system (%
\protect\ref{ac.15}), (\protect\ref{ac.17}) in the compactified variables $X$
and $\protect\eta ~$for $n=-4$ and various values of $\protect\mu_0$.
Equilibrium points at the infinity are marked with red. Point $P_{0}$ is
marked with yellow. Orange points are $P_{1}^{\pm }$, blue points are $%
P_{2}^{\pm }$, purple points are $P_{3}^{\pm }$, and green points are
equilibrium points $P_{4}^{\pm }$. }
\label{fig1a}
\end{figure}

In Figs.~\ref{fig1} and \ref{fig1a}, we present the qualitative evolution of
the phase-space for the two-dimensional system (\ref{ac.15}), (\ref{ac.17})
and the equilibrium points for different values of the free parameters at
the compactified variables. The phase-space portraits can be used to
constraint the initial conditions such that the resulting attractor to
describes the cosmic expansion and acceleration, as described by point $%
P_{4}^{+}$, as also to avoid singularities described by the stationary
points at the infinite regime.

\begin{table}[tbp] \centering%
\caption{Equilibrium points and physical properties for the power-law
potential}%
\begin{tabular}{ccccc}
\hline\hline
\textbf{Point} & \textbf{Existence} & $\mathbf{q}$ & \textbf{Spacetime} & 
\textbf{Stability} \\ \hline
$P_{0}$ & Always & $\nexists $ & Minkowski & Unstable \\ 
$P_{1}^{\pm }$ & Always & $0$ & Scaling & $P_{1}^{-}$ always attractor \\ 
$P_{2}^{\pm }$ & $n\neq -\frac{1}{3}$ & $-\frac{4n}{1+3n}$ & Scaling & $%
P_{2}^{+}$ attractor $n<-\frac{1}{3}$or~$n>0$ \\ 
$P_{3}^{\pm }$ & $\mu _{0}<\frac{1}{2}$ & $-1$ & Not Physical & $P_{3}^{+}$
is an attractor for $n<1$ \\ 
$P_{4}^{\pm }$ & $\frac{2-n}{2-n-4\mu _{0}}>0$ & $-1$ & de Sitter & $%
P_{4}^{+}$ can be attractor \\ 
$P_{0}^{\left( \infty \right) \pm }$ & Always & $\nexists $ & Minkowski & $%
P_{0}^{\left( \infty \right) -}$ is an attractor for $n>0$ \\ 
$P_{1}^{\left( \infty \right) \pm }$ & Always & $-n\infty $ & Big Rip~$%
\left( n>0\right) $/Minkowski~$\left( n<0\right) $ & $P_{1}^{\left( \infty
\right) -}~$ attractor $-\frac{1}{3}<n<0$ \\ 
$P_{-1}^{\left( \infty \right) \pm }$ & Always & $+n\infty $ & Big Crunch~$%
\left( n>0\right) $/Minkowski~$\left( n<0\right) $ & $P_{-1}^{\left( \infty
\right) -}~$ attractor $-\frac{1}{3}<n<0$ \\ \hline\hline
\end{tabular}%
\label{tab2}%
\end{table}%

\subsection{Power-law potential with the cosmological constant}

We consider the introduction of the cosmological constant with the power law
potential. In particular, we consider the potential function $V\left( \phi
\right) =V_{0}\phi ^{n}+2\Lambda $. For this potential, the resulting $%
f\left( R+\mu \mathcal{G}\right) $ theory is 
\begin{equation}
f\left( R+\mu G\right) \simeq \left( R+\mu \mathcal{G}\right) ^{\frac{n}{n-1}%
}-2\Lambda .
\end{equation}%
Moreover, we calculate $\lambda =\frac{V_{0}n\phi ^{n}}{V_{0}\phi
^{n}+2\Lambda }$ and $\Gamma \left( \lambda \right) =\frac{n-1}{\lambda }$,
from where we infer that $\lambda $ for this potential is a varying
variable. Equation (\ref{ac.18}) reads%
\begin{equation}
\frac{d\lambda }{d\tau }=\frac{\lambda }{8}\left( n-\lambda \right) x.
\label{ac.21}
\end{equation}

For this cosmological model, the equilibrium points are defined in a
three-dimensional space with coordinates $A=\left( x\left( A\right) ,\eta
\left( A\right) ,\lambda \left( A\right) \right) $. It follows from (\ref%
{ac.21}) that the stationary points are summarized in two large families $A$
and $B$. The family $A$ consists of points with fixed value $\lambda $, as
given by the algebraic equation $\lambda _{A}\left( \lambda _{A}-n\right) =0$%
, that is $\lambda _{A}=0$~or $\lambda _{A}=n$. On the other hand, family of
points $B$ have $x\left( B\right) =0$ and $\lambda $ is arbitrary.

Specifically, the stationary points are calculated 
\begin{eqnarray*}
A_{0} &=&\left( P_{0},\lambda _{A}\right) ,~A_{1}^{\pm }=\left( P_{1}^{\pm
},\lambda _{A}\right) ,~A_{2}^{\pm }=\left( P_{2}^{\pm },\lambda _{A}\right)
,~ \\
A_{3}^{\pm } &=&\left( P_{3}^{\pm },\lambda _{A}\right) ~\text{and }%
A_{4}^{\pm }=\left( P_{4}^{\pm },\lambda _{A}\right) \text{,}
\end{eqnarray*}%
and%
\begin{equation*}
B_{0}=\left( P_{0},\lambda \right) ,~B_{1}^{\pm }=\left( P_{1}^{\pm
},\lambda \right) ~,~B_{4}^{\pm }=\left( P_{4}^{\pm },\lambda \right) \text{.%
}
\end{equation*}

The physical properties of the asymptotic solutions near the above
equilibrium points are similar to that for the points $P$ examined before,
for $\lambda =\lambda _{A}$. Family of points $A$ recover all the points of
the power-law potential.\ Moreover, the family of points $B$ describe points
where the scalar field $\phi $ is a constant parameter. \ Nevertheless, the
stability properties will be different to that of points $P$. In the
following lines, we discuss the stability properties of the equilibrium
points.

The equilibrium points $A_{0}$,~$A_{1}^{\pm }$,~$A_{4}^{\pm }$,~$B_{0}$,$%
~B_{1}^{\pm }$ and~$B_{4}^{\pm }$ share two eigenvalues with points $P_{0}$
and $P_{1}^{\pm }$ respectively, with the third eigenvalue to be zero.
Consequently, they have the same stability properties as before. We conclude
that $B_{1}^{-}$ and $B_{4}^{+}$ attractor surfaces.

The eigenvalues of the linearized system around the stationary points $%
A_{2}^{\pm }$ are $\lambda _{A}=0:\left\{ \pm 2,\mp 4,\pm 4n\right\} $ and $%
\lambda _{A}=n:\left\{ \mp \frac{2\left( n-1\right) }{1+3n},\mp \frac{4n}{%
1+3n},\mp 4\right\} $. Thus for $\lambda _{A}=0$, points $A_{2}^{\pm }$ are
always saddle points, while for $\lambda _{A}=n$, point $A_{2}^{+}$ is an
attractor for $n<-\frac{1}{3}$ or $n>0$.

Finally, for the equilibrium points $A_{3}^{\pm }$, we calculate the
eigenvalues $\lambda _{A}=0:\frac{1}{\sqrt{1-2\mu _{0}}}\left\{ \mp 4,\mp
1,\pm n\right\} $, and $\lambda _{A}=n:\frac{1}{\sqrt{1-2\mu _{0}}}\left\{
\mp n,\mp \left( 1-n\right) ,\mp 4\right\} $. Hence for $\lambda _{A}=0$,
point $A_{3}^{+}$ is an attractor for $n<0$, while for $\lambda _{A}=n$,
point $A_{3}^{+}$ is an attractor for $0<n<1$. Nevertheless, because points $%
A_{3}^{\pm }$, are not physically accepted, the dynamical system is
well-defined in the region $n>1$.

\subsubsection{Compactified variables}

For the analysis at infinity we employ the compactified variables (\ref%
{ac.22}). Equation (\ref{ac.21}) reads%
\begin{equation}
\frac{d\lambda }{dT}=\frac{\lambda }{8}\left( n-\lambda \right) X,
\end{equation}%
from where we infer that at the infinity, where $X^{2}=1$, stationary points
exist only for $\lambda _{A}=0$ or $\lambda _{A}=n$. The equilibrium points
are 
\begin{equation*}
A_{0}^{\left( \infty \right) \pm }=\left( P_{0}^{\left( \infty \right) \pm
},\lambda _{A}\right) ,~A_{1}^{\left( \infty \right) \pm }=\left(
P_{1}^{\left( \infty \right) \pm },\lambda _{A}\right) ,~A_{-1}^{\left(
\infty \right) \pm }=\left( P_{-1}^{\left( \infty \right) \pm },\lambda
_{A}\right) ,
\end{equation*}%
with the same physical properties as before.

The eigenvalues of the linearized system near points $A_{0}^{\left( \infty
\right) \pm }$ are $\left\{ \pm \frac{1}{4},\pm \frac{1}{8}\left( n-2\lambda
_{A}\right) ,\pm \frac{\lambda _{A}}{8}\right\} $. Therefore, point $%
A_{0}^{\left( \infty \right) -}$ has a stable surface for $\lambda _{A}=0$.,
while the asymptotic solution described by point $A_{0}^{\left( \infty
\right) +}$ is always unstable.

Finally for the equilibrium points $A_{1}^{\left( \infty \right) \pm },~$ $%
A_{-1}^{\left( \infty \right) \pm }$, we derive the eigenvalues $\left\{ \mp 
\frac{\lambda _{A}}{4},\pm \frac{\left( 1+3\lambda _{A}\right) }{4},\pm 
\frac{n-2\lambda }{8}\right\} $. We conclude that~points $A_{1}^{\left(
\infty \right) +},~$ $A_{-1}^{\left( \infty \right) +}$ have a stable
surface for $\lambda _{A}=0$, and they are attractors for $\lambda _{A}=n$,
and $n>0$.

\subsection{Potential $V\left( \protect\phi \right) =V_{0}\left( \protect%
\phi -1\right) ^{n}$}

Now, we consider the potential $V\left( \phi \right) =V_{0}\left( \phi
-1\right) ^{n},~n\neq 0,1$. This potential corresponds to the power-law
theory 
\begin{equation*}
f\left( R+\mu \mathcal{G}\right) \simeq R+\mu \mathcal{G}+\alpha \left(
R+\mu \mathcal{G}\right) ^{\frac{n}{n-1}}.
\end{equation*}

We calculate $\lambda =n\frac{\phi }{\phi -1}$ and $\Gamma \left( \lambda
\right) =1-\frac{1}{n}$. Therefore, the dynamical evolution of parameter $%
\lambda $ is given by equation (\ref{ac.18}), that is,%
\begin{equation}
\frac{d\lambda }{d\tau }=\frac{\lambda }{8}x\left( 1-\frac{\lambda }{n}%
\right) .
\end{equation}

Consequently, the equilibrium points for this cosmological model are the
same with that of power-law potential with the cosmological constant term
examined before.

Similarly, as before, points related to $P_{3}^{\pm }$ can be avoided as
attractor when $n>1$.

At $n=2$, this model modifies the Starobinsky model of inflation~\cite%
{Starobinsky:1980te}.

\section{Introducing Dark Matter}

\label{sec4}

We proceed our study by introducing cold dark matter in the gravitational
model. The introcuction of the dark matter affects the cosmological dynamics
and the cosmological history. It is possible the cosmological model to
describe the matter dominated epoch.

The cold dark matter contributes only to the constraint equation (\ref{ac.11}%
) of the cosmological field equations. Indeed, if $\rho _{m}$ is the energy
component of the cold dark matter, equation (\ref{ac.11}) is modified as%
\begin{equation}
\rho _{m}=6\phi H^{2}+6H\frac{\dot{\phi}}{N}+24H^{3}\mu \frac{\dot{\phi}}{N}%
+V\left( \phi \right) ,  \label{ac.25}
\end{equation}%
where because $\rho _{m}$ is minimally coupled to gravity, it follows 
\begin{equation}
\dot{\rho}_{m}+3H\rho _{m}=0.
\end{equation}

Let $\Omega _{m}=\frac{\rho _{m}}{6\phi H^{2}}$ be the new dimensionless
variable which describes the energy density of the matter component. Then by
using (\ref{ac.25}), the algebraic constraint (\ref{ac.19}) is modified as 
\begin{equation}
\Omega _{m}=y+\frac{3}{4}\eta \left( 8\eta -\frac{x\left( \left( 1-4\mu
\right) \eta ^{2}-1\right) }{1-\eta ^{2}}\right) ,
\end{equation}%
while the rest of the first-order differential equations (\ref{ac.16}), (\ref%
{ac.17}) and (\ref{ac.18}) remain the same. However, due to the existence of
the matter component, now the dynamical system lies in the four-dimensional
space with dynamical variables $\left\{ x,\eta ,y,\lambda \right\} $. 

{The presence of cold dark matter not only adds new stationary points that correspond to $\Omega_m \neq 0$, but it can also make a stable stationary point with $\Omega_m = 0$ unstable one, because $y$ is now an independent variable.}

We consider the power-law potential where $\lambda $ is a constant
parameter. The stationary points are summarized in two families, those where 
$\Omega _{m}=0$, described by the nine points $P_{0},~P_{1}^{\pm }$, $%
P_{2}^{\pm }$, $P_{3}^{\pm }$ and $P_{4}^{\pm }$ found before, and the new
points with $\Omega _{m}\neq 0$; they are~of the form $Q=\left( x\left(
Q\right) ,\eta \left( Q\right) ,y\left( Q\right) \right) $.

The%
\begin{equation*}
Q_{0}^{\pm }=\left( 0,0,\pm 1\right) \,,
\end{equation*}%
with $\Omega _{m}\left( Q_{0}^{\pm }\right) $ describe solutions dominated
by the cold dark matter, that is, $q\left( Q_{0}^{\pm }\right) =\frac{1}{2}$.

The

\begin{equation*}
Q_{1}^{\pm }=\frac{1}{\sqrt{1-2\mu _{0}}}\left( \mp 24,\pm 1,0\right) ,
\end{equation*}%
describe a solution with, $q\left( Q_{1}^{\pm }\right) =-1$, the matter
component has nonzero value $\Omega _{m}\left( Q_{1}^{\pm }\right) =\frac{24%
}{1-2\mu _{0}}$. \ However, similarly to the points $P_{3}^{\pm }$ of
before, $Q_{1}^{\pm }$ are not physical solutions. The requirements, $0\leq
\Omega _{m}\left( Q_{1}^{\pm }\right) \leq 1$ and $1-2\mu _{0}>0$, gives
that $\mu _{0}<-\frac{23}{2}$.

Regarding the stability properties of the equilibrium points, it follows
that $P_{0}$ describes an unstable Minkowski solution. For points $%
P_{1}^{\pm }$ we calculate the eigenvalues $\left\{ \pm 1,\pm 2,\pm
2\right\} $, from where it follows that the $P_{1}^{+}$ is always a source
and $P_{1}^{-}$ is always an attractor.

Moreover, for points $P_{2}^{\pm }$ we calculate the eigenvalues $\left\{
\mp 2,\mp \frac{2\left( 5\lambda -1\right) }{1+3\lambda },\mp \frac{%
7+9\lambda }{1+3\lambda }\right\} $, from where we infer that for $\lambda <-%
\frac{7}{9}$ or $\lambda >\frac{1}{5}$, $P_{2}^{+}$ is an attractor and $%
P_{2}^{-}$ is a source; otherwise for $-\frac{7}{9}<\lambda <\frac{1}{5}$,
the equilibrium points are saddle.

For points $P_{3}^{\pm }$ we derive the eigenvalues $\left\{ \mp \frac{4}{%
\sqrt{1-2\mu _{0}}},\mp \frac{4}{\sqrt{1-2\mu _{0}}},\mp \frac{1-\lambda }{%
\sqrt{1-2\mu _{0}}}\right\} $, that is, for $\lambda <1$, $P_{3}^{+}$ is an
attractor and $P_{3}^{-}$ is a source, otherwise the stationary points are
saddle points. In order points $P_{3}^{\pm }$ not exist, one needs $\mu _{0}>%
\frac{1}{2}$.

\begin{figure}[tbph]
\centering\includegraphics[width=0.5\textwidth]{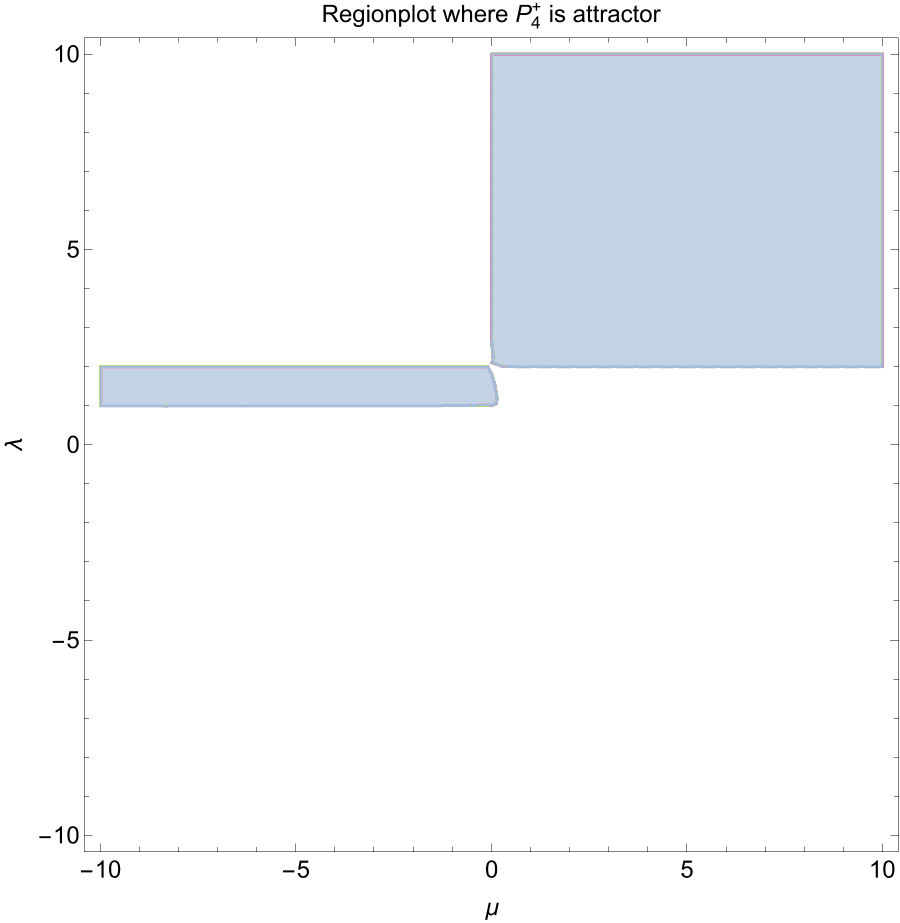}
\caption{The region where point $P_{4}^{+}$ is an attractor. }
\label{fig2}
\end{figure}

The eigenvalues of the linearized system near points $P_{4}^{\pm }$ are
studied numerically within the region $-10$ $\leq \lambda ,\mu_0\leq 10$. We
find that point $P_{4}^{+}$ is an attractor in the region presented in Fig. %
\ref{fig2}.

For points $Q_{0}^{\pm }$ we calculate the eigenvalues, $\left\{ \pm 1,\pm
2,\pm 2\right\} $, that is, $Q_{0}^{+}$ is always a source, and $Q_{0}^{-}$
is always an attractor. Finally, for the linearized system near points $%
Q_{1}^{\pm }$ we derive $\left\{ -\frac{4}{\sqrt{1-2\mu_0}},\frac{4}{\sqrt{%
1-2\mu_0}},\mp \frac{3\left( 1-\lambda \right) }{\sqrt{1-2\mu_0}}\right\} $,
which means that $Q_{1}^{\pm }$ $\ $are always saddle points.

For the analysis at infinity, we employ compactified variables as before.
Thus, the requirement the energy density for the cold dark matter to be
constrained within the region $0\leq \Omega _{m}\leq 1$ reveals that the
only stationary points which are physically accepted are these determined
before. Thus, the stability properties remain unchained.

\section{Conclusions}

\label{sec5}

In this work, we perform a detailed analysis of the phase-space for the
cosmological field equations within the fourth-order $f\left( R,\mathcal{G}%
\right) $-gravity. In particular, we considered spatially flat FLRW
cosmology within the context of $f\left( R,\mathcal{G}\right) =f\left( R+\mu 
\mathcal{G}\right) $ theory of gravity. With the use of the Lagrange
multiplier approach, this theory with a nonlinear function $f$ is equivalent
to a theory with a scalar field nonminimally coupled to the Ricci scalar and
to the\ Gauss-Bonnet term.

We investigate the stability of exact solutions of physical interest
utilizing two methods. Firstly, we considered the effective potential
approach to determine conditions for the stability of the de Sitter
solution. However, this method cannot be used to determine the stability of
scaling solutions or to understand the global dynamics. Therefore, we
introduce dimensionless variables and wrote the cosmological field equations
in equivalent form and get a system of first-order nonlinear differential
equations of these variables.

For the $f\left( R,\mathcal{G}\right) \simeq \left( R+\mu \mathcal{G}\right)
^{\frac{n}{n-1}}$ model, with and without a cosmological constant term, we
have performed a detailed analysis of the stability of the stationary
points. It was found that, in order to avoid the existence of non-physical
solutions parameter $n$ is constrained to be $n>1$. {On the other
hand, the phase-space analysis shows that the points }$P_{3}^{\pm }${%
\ do not appear in the phase-space, since they lead to non-physical
solutions, parameter }$\mu ${\ is constrained to }$\mu >\frac{1}{2}$%
{. Within this framework, the de Sitter solution associated with
point }$P_{4}^{+}${\ together with the scaling solutions
corresponding to point }$P_{2}^{+}${\ can describe a double
inflationary era, with the latter point to be the future attractor. }

Two families of solutions that describe an accelerating expansion can be
related to either two inflationary epochs or to inflation and late-time
acceleration. However, from the study of the phase-space we found that the
majority of the initial conditions leads to a contracting universe. This is
a similar result with that determined recently for the $f\left( \mathcal{G}%
\right) $-cosmology in \cite{Papagiannopoulos:2025uix}. 

Furthermore, we discussed the existence of stationary points for the
power-law model $f\left( R,\mathcal{G}\right) =R+\left( R+\mu \mathcal{G}%
\right) ^{\frac{n}{n-1}}$ which includes the modified Starobinsky theory for 
$n=2$. Again, we found that non-physically solutions are avoided for $n>1$.
Finally, we introduce a dark matter component in the model and perform the
same analysis.

No-ghost conditions for $F(\mathcal{G})$ and $f(\phi ,\mathcal{G})$ models
have been investigated in Refs.~\cite%
{Calcagni:2006ye,DeFelice:2009ak,Tsujikawa:2022aar}. Considering
inhomogeneous linear perturbations, the authors have found no-ghost and
sub-luminal constraint equations. {In this paper, we analyze only
background homogeneous solutions in the spatially flat FLRW metric. We plan
to consider no-ghost conditions for }$f(R+\mu G)${\ model in future
investigations.}

\begin{acknowledgments}
AG was supported by Proyecto Fondecyt Regular 1240247. GL and AP are
grateful for the support of Vicerrector\'{\i}a de Investigaci\'{o}n y
Desarrollo Tecnol\'{o}gico (Vridt) at Universidad Cat\'{o}lica del Norte
through N\'{u}cleo de Investigaci\'{o}n Geometr\'{\i}a Diferencial y
Aplicaciones, Resoluci\'{o}n Vridt No - 096/2022 and Resoluci\'{o}n Vridt No
- 098/2022. GL and AP were economically supported by the Proyecto Fondecyt
Regular 2024, Folio 1240514, Etapa 2025. AP thanks the Universid\'{a}d de la
Frontera \& Universid\'{a}d Austral de Chile for the hospitality provided
when this work was carried out. The research of E.P. and S.V. was conducted
under the state assignment of Lomonosov Moscow State University.
\end{acknowledgments}

\appendix

\section{$f\left( R,\mathcal{G}\right) $ gravity}

\label{app1}

In this Appendix, we introduce the Lagrange multiplier technique in order to
write the gravitational model of $f\left( R,\mathcal{G}\right) $-theory in
an equivalent form of scalar field description.

We assume the $f\left( R,\mathcal{G}\right) $-gravity with Action Integral 
\begin{equation}
S_{f\left( R,\mathcal{G}\right) }=\int d^{4}x\sqrt{-g}f\left( R,\mathcal{G}%
\right) ,  \label{ac.01}
\end{equation}%
where $R$ and $\mathcal{G}$ are the Ricci scalar and the Gauss-Bonnet scalar
respectively of the metric tensor $g_{\mu \nu }$.

In order to derive the field equations, we use Lagrange multipliers to
determine a point-like Lagrangian for the field equations within the
minisuperspace approach.

We introduce the two Lagrange multipliers $\lambda_{1}$ and $\lambda_{2}$.
Hence, the gravitational Action (\ref{ac.01}) reads%
\begin{equation}
S_{f\left( R,\mathcal{G}\right) }=\int d^{4}x\sqrt{-g}\left[ f\left( R,%
\mathcal{G}\right) -\lambda _{1}\left( R-6\left( 2H^{2}+\frac{1}{N}\dot{H}%
\right) \right) -\lambda _{2}\left( \mathcal{G}-24H^{2}\left( H^{2}+\frac{1}{%
N}\dot{H}\right) \right) \right]  \label{ac.02}
\end{equation}%
Variation of the later action integral with respect to the scalars $R$ and $%
\mathcal{G}$ gives $\lambda_{1}=f_{,R}$ and $\lambda_{2}=f_{,\mathcal{G}}$.

We substitute the Lagrange multipliers and the exact expressions (\ref{f.02}%
) and (\ref{f.03}) in the action (\ref{ac.02}), using $\dot{H}=-\frac{\dot{N}%
}{N^{2}}\frac{\dot{a}}{a}+\frac{1}{N}\frac{\ddot{a}}{a}-\frac{1}{N}\left( 
\frac{\dot{a}}{a}\right) ^{2}$, and after integration by parts\footnote{%
We use $\int d^{4}x\left( \frac{6\phi }{N}a^{2}\ddot{a}\right) =\int
d^{4}x\left( -\frac{6a^{2}\dot{a}\dot{\phi}}{N}-\frac{12a\dot{a}^{2}\phi }{N}%
+\frac{6\,a^{2}\dot{a}\dot{N}\phi }{N^{2}}\right) $ and $\int d^{4}x\left( 
\frac{24\dot{a}^{2}\ddot{a}\psi }{N^{3}}\right) =\int d^{4}x\left( -\frac{8%
\dot{a}^{3}\dot{\psi}}{N^{3}}+\frac{24\dot{a}^{3}\dot{N}\psi }{N^{4}}\right) 
$.} it follows%
\begin{equation}
S_{f\left( R,\mathcal{G}\right) }=\int d^{4}xL_{f\left( R,\mathcal{G}\right)
}\left( N,a,\dot{a},\phi ,\dot{\phi},\psi ,\dot{\psi}\right)  \label{ac.03}
\end{equation}%
where now\qquad\ 
\begin{equation}
L_{f\left( R,G\right) }\left( N,a,\dot{a},\phi ,\dot{\phi},\psi ,\dot{\psi}%
\right) ={}-\frac{6}{N}\phi a\dot{a}^{2}-\frac{6}{N}a^{2}\dot{a}\dot{\phi}-%
\frac{8}{N^{3}}\dot{\psi}\dot{a}^{3}-Na^{3}V\left( \phi ,\psi \right) ,
\label{ac.04}
\end{equation}%
and $\phi =f_{,R}$ and $\dot{\psi}=f_{,\mathcal{G}}$, while the scalar field
potential is related to the function $f\left( R,\mathcal{G}\right) $ by the
following expression, 
\begin{equation}
V\left( \phi ,\psi \right) =f_{,R}R+f_{,\mathcal{G}}\mathcal{G}-f\left( R,%
\mathcal{G}\right) ,  \label{ac.05}
\end{equation}%
where $R$ and $\mathcal{G}$ are functions of $\phi $ and $\psi $ now.\qquad

In the scalar field description, the gravitational field equations read%
\begin{equation}
6\phi H^{2}+6H\frac{\dot{\phi}}{N}+24H^{3}\frac{\dot{\psi}}{N}=V\left(\phi
,\psi \right) ,  \label{ac.06}
\end{equation}%
\begin{eqnarray}
0 &=&\frac{2}{N^{2}}\left( 8H\dot{H}\dot{\psi}+\ddot{\phi}+4H^{2}\ddot{\psi}%
\right) +\frac{4}{N}\left( \phi \dot{H}+H\dot{\phi}+4H^{3}\dot{\psi}\right) 
\notag \\
&&-2\frac{\dot{N}}{N^{3}}\left( \dot{\phi}+4H^{2}\dot{\psi}\right) +6\phi
H^{2}-V\left( \phi ,\psi \right)  \label{ac.07}
\end{eqnarray}%
\begin{equation}
V_{,\phi }=6\left( 2H^{2}+\frac{1}{N}\dot{H}\right),  \label{ac.08}
\end{equation}%
\begin{equation}
V_{,\psi }=24H^{2}\left( H^{2}+\frac{1}{N}\dot{H}\right).  \label{ac.09}
\end{equation}%
We remark that the latter two equations are definitions of the scalars.
However, there are less independent second-order differential equations than
the number of the dependent variables, that is, the system is degenerate (or
singular). This means that the two scalar fields $\phi ,~\psi $ are not
dynamically independent.

In the limit where $\phi =const,$ that is, $f\left( R,\mathcal{G}\right)
=R+F\left( \mathcal{G}\right) $, the theory is reduced to that of
Einstein-Gauss-Bonnet model which was discussed previously in Ref.~\cite%
{Papagiannopoulos:2025uix}.

Moreover, if we consider the $f\left( R,\mathcal{G}\right) =f\left( \mathcal{%
X}\right) $ model, with $\mathcal{X}=R+\mu \mathcal{G}$, from the chain rule
for the scalar fields it follows%
\begin{eqnarray}
\phi &=&f_{,R}=\frac{df}{d\mathcal{X}}\frac{d\mathcal{X}}{dR}=\frac{df}{d%
\mathcal{X}}, \\
\psi &=&f_{,\mathcal{G}}=\frac{df}{d\mathcal{X}}\frac{d\mathcal{X}}{dG}=\mu 
\frac{df}{d\mathcal{X}}=\mu \phi .
\end{eqnarray}

Hence, we end with the single scalar field point-like Lagrangian function%
\begin{equation}
L_{f\left( R,\mathcal{G}\right) }^{A}\left( N,a,\dot{a},\phi ,\dot{\phi}%
\right) ={}-\frac{6}{N}\phi a\dot{a}^{2}-\frac{6}{N}a^{2}\dot{a}\dot{\phi}-%
\frac{8\mu }{N^{3}}\dot{\phi}\dot{a}^{3}-Na^{3}V\left( \phi \right) .
\label{ac.10a}
\end{equation}%
where now 
\begin{equation}
V\left( \phi \right) =\mathcal{X}F_{,\mathcal{X}}\left( \mathcal{X}%
\right)-F\left( \mathcal{X}\right) \,.
\end{equation}

\end{document}